\documentclass[12pt]{iopart}

\usepackage{graphicx}

\def\be{\begin{equation}}
\def\te{\end{equation}}
\def\ee{\end{equation}}
\def\ba{\begin{eqnarray}}
\def\bea{\begin{eqnarray}}
\def\nn{\nonumber\\}
\def\tea{\end{eqnarray}}
\def\ea{\end{eqnarray}}
\def\eea{\end{eqnarray}}

\begin{document}
\title{Wave propagation in non Gaussian random media}

\author{Mariano Franco$^1$ and Esteban Calzetta$^{2}$}
\address{$^1$ Physics Department, Buenos Aires University, Ciudad Universitaria, Pabell\'on I, Buenos Aires, 1428, Argentina and IAFE-CONICET}
\address{$^2$ Physics Department, Buenos Aires University, Ciudad Universitaria, Pabell\'on I, Buenos Aires, 1428, Argentina and IFIBA-CONICET}
\eads{\mailto{mfranco@df.uba.ar}, \mailto{calzetta@df.uba.ar}}

\begin{abstract}
We develop a compact perturbative series for accoustic wave propagation in a medium with a non Gaussian stochastic speed of sound. We use Martin - Siggia and Rose auxiliary field techniques to render the classical wave propagation problem into a ``quantum'' field theory one, and then frame this problem within so-called Schwinger - Keldysh of closed time-path (CTP) formalism.  Variation of the  so-called two-particle irreducible (2PI) effective action (EA), whose arguments are both the mean fields and the irreducible two point correlations, yields the Schwinger-Dyson and the Bethe-Salpeter equations. We work out the loop expansion of the 2PI CTP EA and show that, in the paradigmatic problem of overlapping spherical intrusions in an otherwise homogeneous medium, non Gaussian corrections may be much larger than Gaussian ones at the same order of loops
\end{abstract}

\section{Introduction}
Wave propagation in a heterogeneous medium \cite{TsangKong01,Furutsu82,AkkMon07,Voronovich94,Sheng06} is a problem with an enormous range of aplications, from remote sensing to Anderson localization \cite{SaFeMa12,ElaZyl06,TigSki03}. In many applications, the medium may be modelled as having a point-dependent, partially random speed of sound (in this paper we shall discuss the scalar wave equation only; the generalization of our work to the electromagnetic case is straightforward).  There are two main objects to be computed, the self energy or mass operator, which is essentially the inverse square  of the speed of propagation of the wave in the medium and thus necessary to compute the coherent part of the field, and the intensity operator, which is necessary to compute field fluctuations and thereby the incoherent scattering. We give precise definitions of both these objects below. These objects are constrained by the symmetries of the theory, namely flux conservation (in the scalar case, in the electromagnetic case the relevant conserved quantity is energy) and reciprocity \cite{Potton04,TigMay98,Hovenier68}. Flux conservation implies a constraint which mass and intensity operator must satisfy, the so called Ward identity \cite{NiChSh01,BarOzr01,NiChSh98}.

In $d=3$ dimensions, the problem is already too hard for a closed form solution, and one must resort to perturbation theory, the small parameter being associated to the strenght of random fluctuations in the speed of sound. Field theory methods have been extensively applied in the development of suitable perturbative schemes \cite{Frisch68,WolVol82,VolWol92}. In the most straightforward formulation, the field equation is first transformed into an integral equation which is then solved iteratively. One obtains in this way a linear mean field equation for the coherent field, a Schwinger - Dyson equation for the Green function associated to the mean field equation (which measures the response of the mean field to an external source) and the Bethe - Salpeter equation for the two-point correlator of incoherent field fluctuations \cite{SalBet51,BetSal57,ItzZub80,Furutsu85,SouBer03a,SouBer03b,SouBer03c}. One may write down these equations for a given realization of the stochastic speed of sound. They will contain kernels composed of products of the random part of the inverse square speed of sound at various points. Replacing those products by their statistical averages one finally finds the mass and intensity operators, and therefore the final equations for the Green function and the incoherent field correlations. The first nontrivial approximation yields the so-called nonlinear approximation for the mass operator and the ladder approximation for the intensity operator (see below). These approximations are consistent with the Ward identity but not with reciprocity. To restore reciprocity (and also to account for other phenomena such as coherent backscattering and Anderson localization) it is necessary to move to higher orders in perturbation theory \cite{Knothe12,KnoWel13,KCLGMM12,EcBuWe12,FABSAMM,Slavov06,RosNie99,TiWiLa95,HaStBa94,MaTsIs90,VolWol80} (see also \cite{IIR11,IIR12a,IIR12b,IIR12c}).

In doing so, one encounters expressions involving products of the random inverse square speed of sound at three or more different points. The usual perturbative approach computes these higher correlations as if the random  fluctuations were Gaussian, namely, higher correlations are factorized into binary correlations according to Wick's theorem. The only information kept about the medium is the two point correlation of fluctuations. This is unsatisfactory as many relevant applications involve non Gaussian fluctuations \cite{EmeOrt11}; for example, the fluctuating speed of sound which is obtained by randomly distributing bubbles into an otherwise homogeneous medium \cite{Bruno91,Torquato02} is generally non Gaussian, as we will show below. 

Hori and coworkers have developed a statistical theory which may deal with non Gaussian fluctuations \cite{Hori75a,Hori75b,Hori75c,Hori77,Hori73b,Hori73a,Hori74}. However, to the best of our knowledge, their work is oriented mostly to computing the mass operator. Our aim is to develop a formalism which is more suitable to the computation of both coherent and incoherent scattering, and in the process to present perturbation theory in a way which we believe is more compact and therefore easier to pursue to higher orders. As a particularly compelling feature, the perturbative scheme to be presented below has flux conservation built in order by order in perturbation theory.

To achieve our goals we shall use Martin, Siggia and Rose auxiliary field techniques \cite{MaSiRo73,DomPel78,CLL89,Kamenev,ZanCal02}
 to render the classical wave propagation problem into a ``quantum'' field theory one, and then frame this problem within the so-called Schwinger - Keldysh or closed time-path (CTP) formalism \cite{CalHu08,Schw61,Mahan62,BakMah63a,BakMah63b,Kel64}. By coupling the different fields and their binary products to fictitious forces, one obtains a CTP generating function for the field correlators of any order. The Legendre transform of the CTP generating function yields the so-called two-particle irreducible (2PI) effective action (EA), whose arguments are both the mean fields and the correlators \cite{LutWar60,DomMar64a,DomMar64b,CoJaTo74}. Variation of the effective action yields the Schwinger-Dyson and the Bethe-Salpeter equations, whereby it is straightforward to identify the mass and intensity operators.

The key to the power of the method is that the seeming complexity of dealing with a larger number of fields (auxiliary and ghost \cite {NegOrl98,Cal04}
fields on top of the physical ones) is overridden by the fact that the 2PI EA is built from Feynman graphs with no external legs and which are two-particle irreducible, meaning that they remain connected after cutting any two internal legs. The number of these graphs at any finite order in perturbation theory is small enough that the computational effort stays manageable, and their structure is so tightly constrained that it is possible to provide proofs of key features, such as flux conservation, at any order \cite{PRC09}.

The rest of the paper is organized as follows. In next section we give a brief overview of the perturbative theory based on a direct iteration of the field equations. This allows us to give precise definitions of the key concepts, such as Green functions, two-point correlations, Schwinger-Dyson and Bethe-Salpeter equations and the Ward identity. In the following  section, we present our approach, based on the Martin - Siggia - Rose and CTP formalisms. In this section we work \emph{as if} we had a closed-form expression for the 2PI EA. In the following section IV we discuss the loop expansion of the 2PI EA. Then, as an application, we discuss the sensibility of the mass and intensity operators to the third order correlations in the inverse square speed of sound in the case of a medium composed of spherical, interpenetrating bubbles randomly dispersed into a homogeneous matrix.

We conclude with some brief final remarks.

\section{The direct approach}
We consider a complex scalar field $\Phi$ obeying a wave equation

\be
\left[ \mathbf{\Delta}+\epsilon \left( x\right)  \omega^2\right]  \Phi=-j
\label{1}
\te
where $\epsilon =1/c^ 2 $, $c$ being the speed of sound in the medium. In turn 

\be 
\epsilon \left( x\right)=\bar\epsilon \left( x\right)+\varepsilon \left( x\right)
\te 
$\varepsilon$ is a stochastic (real) variable with zero mean. We consequently split $\Phi=\phi+\varphi$, where $\left\langle \varphi\right\rangle=0$. 

There are a number of properties of the theory which can derived directly from the equations of motion, independently of any perturbative scheme. We begin by reviewing some of them.

Observe that for a given realization of the noise, eq. (\ref{1}) is linear and admits a Green function

\be
\left[ \mathbf{\Delta}+\epsilon \left( x\right) \omega^2\right]G_{\epsilon}\left(x,y\right)=-\delta\left(x-y\right)
\label{5}
\te
The background field $\phi$ does not appear in this equation, and so $G_{\epsilon}$ is background field independent. The solution to eq. (\ref{1}) is

\be
\Phi\left(x\right)=\int\!dy\:G_{\epsilon}\left(x,y\right)j\left(y\right)
\label{6}
\te
Taking the expectation value of eq. (\ref{6}) we get

\be
\phi\left(x\right)=\int\!dy\:G\left(x,y\right)j\left(y\right)
\label{7}
\te
where

\be
G\left(x,y\right)=\left\langle G_{\epsilon}\left(x,y\right)\right\rangle
\label{8}
\te
is independent of $\phi$. Taking the expectation value of eq. (\ref{5}) we get

\be
\left[ \mathbf{\Delta}+\bar\epsilon \left( x\right) \omega^2\right]G\left(x,y\right)=-\delta\left(x-y\right)- \omega^2\left\langle \varepsilon\left(x\right)G_{\epsilon}\left(x,y\right)\right\rangle
\label{9}
\te
We \emph{define} the self energy $\Sigma\left(x,y\right)$ from the identity

\be
 \omega^2\left\langle \varepsilon\left(x\right)G_{\epsilon}\left(x,y\right)\right\rangle=\int\!dz\:\Sigma\left(x,z\right)G\left(z,y\right)
\label{10}
\te
Therefore

\be
\left[\mathbf{\Delta}+\bar\epsilon \left( x\right) \omega^2\right]G\left(x,y\right)+\int\!dz\:\Sigma\left(x,z\right)G\left(z,y\right)
=-\delta\left(x-y\right)
\label{11}
\te
and also the mass operator $Q\left(x,y\right)$ from 

\be
\omega^2\left\langle \varepsilon\left(x\right)\varphi\left(x\right)\right\rangle=\int\!dz\:Q\left(x,z\right)\phi\left(z\right)
\label{14}
\te
Therefore the mean field equation reads
\be
\left[\mathbf{\Delta}+\bar\epsilon \left( x\right) \omega^2\right]\phi\left(x\right)+\int\!dz\:Q\left(x,z\right)\phi\left(z\right)=-j\left(x\right)
\label{12}
\te
On the other hand, acting with $\mathbf{\Delta}+\bar\epsilon \left( x\right)\omega^2+\Sigma$ on both sides of eq. (\ref{7}) we conclude that 

\be
\left[\mathbf{\Delta}+\bar\epsilon \left( x\right) \omega^2\right]\phi\left(x\right)+\int\!dz\:\Sigma\left(x,z\right)\phi\left(z\right)=-j\left(x\right)
\label{12b}
\te
Therefore $Q=\Sigma$. Provided  natural boundary conditions are chosen, $G_{\epsilon}$ is symmetric. This is the so-called reciprocity property.

From the field $\Phi$ we may construct the current

\be 
J_{\epsilon}=\left( -i\right) \left[ \Phi^*\nabla\Phi-\Phi\nabla\Phi^*\right] 
\te 
which obeys 

\be 
\nabla J_{\epsilon}=\left( -i\right) \left[-\Phi^* j+\Phi j^*\right] 
\te 
Therefore the expectation value 

\be 
J=\left\langle J_{\epsilon}\right\rangle 
\te 
conserves charge in the mean, namely,

\be
\nabla J=\left( -i\right) \left[-j\phi^*+j^*\phi\right] 
\label{17}
\te
To compute the current we  need the expectation value

\be
\left\langle \Phi^*\left(x'\right)\Phi\left(x\right)\right\rangle=\phi^*\left(x'\right)\phi\left(x\right)+\left\langle \varphi^*\left(x'\right)\varphi\left(x\right)\right\rangle
\label{15}
\te
We define the ``intensity'' operator from the identity (the so-called Bethe-Salpeter equation)
\be
\left\langle \varphi^*\left(x'\right)\varphi\left(x\right)\right\rangle=\int\:dzdz'dydy'\:G\left(x,z\right)
G^*\left(x',z'\right)I\left[z,y;z',y'\right]\left\langle \Phi^*\left(y'\right)\Phi\left(y\right)\right\rangle
\label{16}
\te
The average of the conservation law over space-time yields the Ward identity

\be
Q\left(y,y'\right)^*-Q\left(y',y\right)-\int\:dzdz'\left[G^*\left(z,z'\right)-G\left(z',z\right)\right]I\left[z,y;z',y'\right]=0
\label{20}
\te

The simplest expression for the mass operator is given by the so-called  nonlinear approximation 

\be
\Sigma_{nonlin}\left(x,x'\right)=Q_{nonlin}\left(x,x'\right)=\omega^4C\left(x,x'\right)G\left(x,x'\right)
\label{nonlinear}
\te
Once this is accepted, the ladder approximation 

\be
I_{ladder}\left[x,y;x',y'\right]=\omega^4C\left(x,x'\right)\delta\left(x-y\right)\delta\left(x'-y'\right)
\label{ladder}
\te
provides the simplest solution to the Ward identity.

We may also write

\be 
\left\langle \Phi^*\left(x'\right)\Phi\left(x\right)\right\rangle=\int\:dydy'\;U\left[ x,y;x',y'\right] j\left( y\right) j^*\left( y'\right) 
\te 
Then 

\be 
U\left[ x,y;x',y'\right]=\left\langle G_{\epsilon}\left( x,y\right) G_{\epsilon}^*\left( x',y'\right) \right\rangle 
\te 
The symmetry of the propagators implies the reciprocity condition 

\be 
U\left[ x,y;x',y'\right]=U\left[ x,y;y',x'\right]
\te
On the other hand we have the identity

\be 
U=\left[ 1 -GG^*I\right] ^{-1}GG^*
\label{iter}
\te
The ladder approximation, namely, the value of $U$ which is obtained if we substitute $I_{ladder}$ into eq. (\ref{iter}) violates the reciprocity condition. 

\section{The functional approach}
The functional approach begins with the observation that moments of the stochastic $\Phi$ field may be derived from a generating functional

\be
e^{iW\left[\mathcal{J},\mathcal{J}^* \right] }=\int\:D\Phi D\Phi^*\:\mathcal{P}\left[\Phi,\Phi^*\right]\:e^{i\left[\mathcal{J}\Phi+\mathcal{J}^*\Phi^*\right]}
\label{21}
\te
Integration is understood in the exponent. $\mathcal{P}$ is the probability density

\be
\mathcal{P}\left[\Phi,\Phi^*\right]=\int\:D\varepsilon\:\mathcal{F}\left[\varepsilon\right]\:\delta\left(\Phi-\Phi\left[\epsilon,j\right]\right)
\delta\left(\Phi^*-\Phi\left[\epsilon,j^*\right]\right)
\label{22}
\te
where $\mathcal{F}\left[\varepsilon\right]$ is the probability density for the multiplicative noise
and $\Phi\left[\epsilon,j\right]$ is the solution to eq. (\ref{1}) for a given noise realization. Observe that

\be
\delta\left(\Phi-\Phi\left[\epsilon,j\right]\right)=
\mathrm{Det}\left[\mathbf{D_{\epsilon}}\right]\delta\left(\mathbf{D_{\epsilon}}\Phi+j\right)
\label{23}
\te
where $\mathbf{D_{\epsilon}}=\mathbf{\Delta}+\epsilon \omega^2$. We exponentiate the delta functions by adding auxiliary fields $\psi$ and $\psi^*$ and the determinants by adding ghost fields $\left(\xi,\eta,\zeta,\vartheta\right)$ \cite {NegOrl98,Cal04} . We also parameterize

\be
\mathcal{F}\left[\varepsilon\right]\propto\exp\left\{-F\left[\varepsilon\right]\right\}
\label{22b}
\te

\be 
F\left[\varepsilon\right]=\sum_{n=0}^{\infty}\frac{1}{n!}\int\;dx_1...dx_n\;f_n\left( x_1,...,x_n\right) \varepsilon\left( x_1\right)... \varepsilon\left( x_n\right) 
\label{Taylor}
\te 
$f_0$ is an overall normalization constant which may be absorbed into the path integration measure; $f_1$ is included to enforce $\left\langle \varepsilon\right\rangle =0$. Observe that in practice we have traded the non gaussian statistics by an infinite hierarchy of interactions \cite{CalHu88,ZSHY80,CSHY85}; this is manageable, nevertheless, because only a finite number of interactions will be effective at any order in the loop expansion \cite{CalHu95a}

The result is that the original stochastic theory is equivalent to a quantum field theory for fields 

\be
X^A=\left(\varepsilon,\Phi,\psi^*,\Phi^*,\psi,\xi,\eta,\zeta,\vartheta\right)
\label{24}
\te
($A=0-8$) with classical action 

\be
S= iF\left[\varepsilon\right]+\psi^*\left[\mathbf{D_{\epsilon}}\Phi+j\right]+\psi\left[\mathbf{D_{\epsilon}}\Phi^*+j^*\right]
+i\xi\mathbf{D_{\epsilon}}\eta+i\zeta\mathbf{D_{\epsilon}}\vartheta
\label{25}
\te
Observe that $\eta$ and $\vartheta$ have ghost charge $1$, $\xi$ and $\zeta$ have ghost charge $-1$ and all other fields have ghost charge $0$. For now on we shall refer to $\Phi,\psi^*,\Phi^*$ and $\psi$ as ``matter'' fields, and to $\xi,\eta,\zeta$ and $\vartheta$ as ghost fields.

The structure of the classical action has a number of consequences for the expectation value. Foremost, observe that $W\left[\mathcal{J}=\mathcal{J}^*=0\right]=0$ independently of $j$ and $j^*$. Therefore we have that \emph {the expectation value of any product of $\psi^*$ and $\psi$ fields vanishes}, since any such product can be obtained as some derivative of $W$ with respect to $j$ and $j^*$. With the same argument we may show that 

\be
\left\langle \psi^*\epsilon\right\rangle=\left\langle \psi\epsilon\right\rangle=\left\langle \psi^*\varphi^*\right\rangle=\left\langle \psi\varphi\right\rangle=0
\label{25.7}
\te
The identity

\be
\int\:DX^A\:\frac{\delta S}{\delta\psi^*}\:e^{iS}=0
\label{25.1}
\te
implies that

\be
\mathbf{D}_0\phi+\omega^2\left\langle \varepsilon\varphi\right\rangle=-j
\label{25.2}
\te
where 

\be 
\mathbf{D}_0=\mathbf{\Delta}+\bar\epsilon \omega^2
\te 
Thus reproducing eq. (\ref{12b}), after we identify $\phi$ as the expectation value of the ``quantum'' field $\Phi=\phi+\varphi$, We therefore get an interpretation of the mass operator $Q$, now defined through eq. (\ref{14})

\be
Q=\omega^ 2\frac{\delta\left\langle \varepsilon\varphi\right\rangle}{\delta\phi}
\label{25.5}
\te
The propagator is defined from eq. (\ref{7})

\be
G\left(x,x'\right)=\frac{\delta\phi\left(x\right)}{\delta j\left(x'\right)}
\te
The variational derivative may be computed from the path integral to get

\be
G=\frac{\delta\phi}{\delta j}=i\left\langle \varphi\psi^*\right\rangle
\label{25.6}
\te
\subsection{The 2-particle irreducible effective action}
The most convenient way to derive the equations of motion for mean field and propagators is through the 2PI EA. We are interested in a situation where only $\Phi$ and $\Phi^*$ develop a nontrivial expectation value. In particular, all objects with nonzero ghost number will vanish on shell.

To obtain the 2PI EA we first generalize the generating functional Eq. (\ref{21}) by including sources for all fields, and also sources coupled to binary products of fields

\be
e^{iW\left[\mathcal{J}_A,\mathcal{K}_{AB} \right] }=\int\:DX^A\:e^{i\left[S\left[ X^A\right] +\mathcal{J}_AX^A+\frac12X^A\mathcal{K}_{AB}X^ B\right]}
\label{21_2PI}
\te
This generating functional generates the expectation values and binary correlations of the fields

\bea 
W\overleftarrow{\frac{\delta}{\delta\mathcal{J}_A}}&=&\bar{X}^A \nn
W\overleftarrow{\frac{\delta}{\delta\mathcal{K}_{AB}}}&=&\frac12 \theta^{BA}\theta^B\left[ \bar{X}^A\bar{X}^B+G^{AB}\right] 
\tea
where $\bar{X}^A=\left\langle X^A\right\rangle$, $G^{AB}=\left\langle X^AX^B\right\rangle-\bar{X}^A\bar{X}^B$, and $\theta^B=\left( -1\right) ^{q_B}$, $\theta^{BA}=\left( -1\right) ^{q_Bq_A}$, $q_B$ being the ghost charge of field $X^B$. Therefore, if we define the Legendre transform \cite{CalHu08,Cal04}

\be 
\Gamma =W-\bar{X}^A\mathcal{J}_A-\frac12 \bar{X}^A\mathcal{K}_{AB}\bar{X}^B-\frac12 \theta^{BA}\theta^BG^{AB}\mathcal{K}_{AB}
\label{g2pi}
\te 
variation of the EA yields the equations of motion

\bea 
\frac{\delta}{\delta\bar{X}^A}\Gamma &=&-\mathcal{J}_A-\mathcal{K}_{AB}\bar{X}^B\nn
\frac{\delta}{\delta G^{AB}}\Gamma &=&-\frac12 \theta^{BA}\theta^B\mathcal{K}_{AB}
\label{2pieqs}
\tea
We make the ansatz

\be 
\Gamma =S\left[  \bar{X}^A\right] +\frac12 \theta^{BA}\theta^BG^{AB}S_{,AB}-\frac i2\ln\; \mathrm{sdet}\left[ G^{AB}\right] +\Gamma_2\left[ G^{AB}\right] +\mathrm{const.}
\te 
$S_{,AB}$ comes from the second variation of the action

\be 
S\left[  \bar{X}^A+x^A\right]=S\left[  \bar{X}^A\right]+S_{,A}x^A+\frac12x^AS_{,AB}x^B+S_{int}\left[ x^A\right] 
\te 
In our case $S_{int}$ is independent of the $ \bar{X}^A$; this will be a large simplification in what follows. The so called superdeterminant $ \mathrm{sdet}$ is defined through the Gaussian integration formula

\be 
\left[\mathrm{sdet}\left[ G^{AB}\right]\right] ^{1/2}=\int\;Dx^A\;e^{\frac{-1}2x^A\left(G^{-1} \right)^{\left(L \right) }  _{AB}x^B} 
\te 
we have absorbed all universal constants into the measure, and the superscript $L$ denotes a left inverse. We have the variational formula 

\be
\frac{\delta}{\delta G^{AB}}\left[\mathrm{sdet}\left[ G^{AB}\right]\right]=\theta^{BA}\theta^A\left[\mathrm{sdet}\left[ G^{AB}\right]\right]\left( G^{-1}\right)^{\left( R\right) } _{AB}
\te
The superscript $R$ reminds us that this is the right inverse of $G^{AB}$. 
Then $\Gamma_2\left[ G^{AB}\right]$ is given by

\be 
e^{i\Gamma_2}=\left[\mathrm{sdet}\left[ G^{AB}\right]\right] ^{-1/2}\int\;Dx^A\;e^{\left\lbrace \frac{-1}2x^A\left( G^{-1}\right)^{\left(L \right) }  _{AB}x^B+i\left( S_{int}\left[ x^A\right] +{\chi}_Ax^A+\frac 12x^A\kappa_{AB}x^B\right) \right\rbrace } 
\te 
The external sources enforce the constraints $\left\langle x^A\right\rangle =0$, $\left\langle x^Ax^ B\right\rangle =G^{AB}$. Because of these constraints, the sum of tadpole insertions and of self energy corrections to any given graph vanishes; this includes all graphs where the sources appear explicitly. Therefore to compute $\Gamma_2$ we may disregard the external sources, provided at the same time we disregard all graphs which are one or two particle reducible, that is, that may be seen as a tadpole insertion or a self energy correction upon some other graph. In other words, $\Gamma_2$ is the sum of all the vacuum, two-particle irreducible Feynman graphs generated from the vertices contained in $S_{int}$ carrying propagators $G^{AB}$ in their internal lines.

The equations for the propagators are

\be
S_{,AB}-i\left( G^{-1}\right) ^{\left(L \right) }_{AB}+2\theta^{AB}\theta^{B}\frac{\delta\Gamma_2}{\delta G^{AB}}=0
\label{26}
\te
Now

\be
S_{,AB}=\left(\begin{array}{cc}
H_{ab} & 0 \\
0      & M^{ij}
\end{array}
\right)
\label{27}
\te

\be
H_{,ab}=\left(\begin{array}{ccccc}
if_2& 0            & \omega^2\phi        & 0            & \omega^2\phi^*     \\
0      & 0            & \mathbf{D_0} & 0            &    0        \\
\omega^2\phi  & \mathbf{D_0} & 0            & 0            &    0        \\
0      & 0            & 0            & 0            &\mathbf{D_0} \\ 
\omega^2\phi^*& 0            & 0            & \mathbf{D_0} & 0 
\end{array}
\right)
\label{28}
\te

\be
M^{ij}=\left(\begin{array}{cccc}
0              & i\mathbf{D_0} & 0              &      0        \\
-i\mathbf{D_0} & 0             & 0              &      0        \\
0              & 0             & 0              & i\mathbf{D_0} \\ 
0              & 0             & -i\mathbf{D_0} &      0 
\end{array}
\right)
\label{29}
\te
Let us investigate the bosonic propagators. We have

\be
S_{,rt}G^{ts}=i\delta^s_r-2\frac{\delta\Gamma_2}{\delta G^{rt}}G^{ts}
\label{30}
\te
Setting $r=1$ ($X^1=\varphi$) we get

\be
\mathbf{D}_0\left\langle \psi^* X^s\right\rangle=i \delta^s_1-2\frac{\delta\Gamma_2}{\delta \left\langle \varphi X^t\right\rangle}\left\langle X^tX^s\right\rangle
\label{31}
\te
If we further set $s=2$ and $s=4$ we get

\be
\frac{\delta\Gamma_2}{\delta \left\langle \varphi \varphi\right\rangle}=\frac{\delta\Gamma_2}{\delta \left\langle \varphi \varphi^*\right\rangle}=0
\label{31.1}
\te
Similarly, from $r=3$

\be
\mathbf{D}_0\left\langle \psi X^s\right\rangle=i \delta^s_3-2\frac{\delta\Gamma_2}{\delta \left\langle \varphi^* X^t\right\rangle}\left\langle X^tX^s\right\rangle
\label{32}
\te
Setting $s=4$ we get

\be
\frac{\delta\Gamma_2}{\delta \left\langle \varphi^* \varphi^*\right\rangle}=0
\label{33}
\te
Now setting $s=0$ we get

\be
\frac{\delta\Gamma_2}{\delta \left\langle \varepsilon \varphi\right\rangle}=\frac{\delta\Gamma_2}{\delta \left\langle \varepsilon \varphi^*\right\rangle}=0
\label{34}
\te
We see that eq. (\ref{31}) reduces to

\be
\left[\mathbf{D}_0+2\frac{\delta\Gamma_2}{\delta \left\langle \varphi \psi^*\right\rangle}\right]\left\langle \psi^* \varphi\right\rangle=i \mathbf{1}
\label{35}
\te
Let us go back to eq. (\ref{30}) and write the remaining equations. With $r=0$ we get

\bea
\left[if_2+2\frac{\delta\Gamma_2}{\delta \left\langle \varepsilon \varepsilon\right\rangle}\right]\left\langle \varepsilon X^s\right\rangle&+&\left[\omega^2\phi+2\frac{\delta\Gamma_2}{\delta \left\langle \varepsilon \psi^*\right\rangle}\right]\left\langle \psi^* X^s\right\rangle\nn
&+&\left[\omega^2\phi^*+2\frac{\delta\Gamma_2}{\delta \left\langle \varepsilon \psi\right\rangle}\right]\left\langle \psi X^s\right\rangle=i\delta^s_0
\label{37}
\tea
With $s=0$ we get

\be
\left[if_2+2\frac{\delta\Gamma_2}{\delta \left\langle \varepsilon \varepsilon\right\rangle}\right]\left\langle \varepsilon \varepsilon\right\rangle=i\mathbf{1}
\label{38}
\te
We shall write the $\varepsilon$  correlation $\left\langle \varepsilon \left( x\right) \varepsilon\left( x'\right) \right\rangle=C\left( x,x'\right) $ for short. With $s=1$

\be
\left\langle \varepsilon\varphi\right\rangle=C\left[\omega^2\phi+2\frac{\delta\Gamma_2}{\delta \left\langle \varepsilon \psi^*\right\rangle}\right]G'
\label{39}
\te
where  $G'=i\left\langle \psi^*\varphi\right\rangle$. Now set $r=2$

\bea
&&\left[\omega^2\phi+2\frac{\delta\Gamma_2}{\delta \left\langle \psi^*\varepsilon\right\rangle}\right]\left\langle \varepsilon X^s\right\rangle +\left[\mathbf{D}_0+2\frac{\delta\Gamma_2}{\delta \left\langle \psi^*\varphi \right\rangle}\right]\left\langle\varphi X^s\right\rangle\nn&+&2\frac{\delta\Gamma_2}{\delta \left\langle \psi^*\psi^* \right\rangle}\left\langle\psi^* X^s\right\rangle+2\frac{\delta\Gamma_2}{\delta \left\langle \psi^*\varphi^* \right\rangle}\left\langle\varphi^* X^s\right\rangle+2\frac{\delta\Gamma_2}{\delta \left\langle \psi^*\psi \right\rangle}\left\langle\psi X^s\right\rangle=i\delta^s_2
\label{40}
\tea
With $s=2$ this leads to 

\bea
&&\left[\omega^2\phi+2\frac{\delta\Gamma_2}{\delta \left\langle \psi^*\varepsilon\right\rangle}\right]\left\langle \varepsilon \psi^*\right\rangle +\left[\mathbf{D}_0+2\frac{\delta\Gamma_2}{\delta \left\langle \psi^*\varphi \right\rangle}\right]\left\langle\varphi \psi^*\right\rangle\nn&+&2\frac{\delta\Gamma_2}{\delta \left\langle \psi^*\psi^* \right\rangle}\left\langle\psi^* \psi^*\right\rangle+2\frac{\delta\Gamma_2}{\delta \left\langle \psi^*\varphi^* \right\rangle}\left\langle\varphi^* \psi^*\right\rangle+2\frac{\delta\Gamma_2}{\delta \left\langle \psi^*\psi \right\rangle}\left\langle\psi \psi^*\right\rangle=i\mathbf{1}
\label{40b}
\tea
Now $\left\langle \varepsilon \psi^*\right\rangle=\left\langle\psi^* \psi^*\right\rangle=\left\langle\psi \psi^*\right\rangle=\left\langle\varphi^* \psi^*\right\rangle=0$, so this reduces to

\be 
\left[\mathbf{D}_0+2\frac{\delta\Gamma_2}{\delta \left\langle \psi^*\varphi \right\rangle}\right]\left\langle\varphi \psi^*\right\rangle=i\mathbf{1}
\label{40c}
\te
We  identify $G=i\left\langle \varphi\psi^*\right\rangle$ and therefore

\be
\Sigma =Q =2\frac{\delta\Gamma_2}{\delta \left\langle \psi^*\varphi \right\rangle}
\label{41}
\te
With $s\neq 2$ we get

\bea
\left\langle \varphi X^s\right\rangle&=&G\left\{\left[\omega^2\phi+2\frac{\delta\Gamma_2}{\delta \left\langle \psi^*\varepsilon\right\rangle}\right]\left\langle \varepsilon X^s\right\rangle\right.\nn
&+&\left.2\frac{\delta\Gamma_2}{\delta \left\langle \psi^*\psi^* \right\rangle}\left\langle\psi^* X^s\right\rangle+2\frac{\delta\Gamma_2}{\delta \left\langle \psi^*\varphi^* \right\rangle}\left\langle\varphi^* X^s\right\rangle+2\frac{\delta\Gamma_2}{\delta \left\langle \psi^*\psi \right\rangle}\left\langle\psi X^s\right\rangle\right\}
\label{42}
\tea
If $s=0$

\be
\left\langle \varphi \varepsilon\right\rangle=G\left\{\left[\omega^2\phi+2\frac{\delta\Gamma_2}{\delta \left\langle \psi^*\varepsilon\right\rangle}\right]C
+2\frac{\delta\Gamma_2}{\delta \left\langle \psi^*\varphi^* \right\rangle}\left\langle\varphi^* \varepsilon\right\rangle\right\}
\label{43}
\te
Compared with eq. (\ref{39}) we conclude that

\be
\frac{\delta\Gamma_2}{\delta \left\langle \psi^*\varphi^* \right\rangle}=0
\label{44}
\te
We may reach the same conclusion by setting $s=4$. Finally, with $s=1$ and $s=3$ we get

\be
\left\langle \varphi \varphi\right\rangle =G\left\{\left[\omega^2\phi+2\frac{\delta\Gamma_2}{\delta \left\langle \psi^*\varepsilon\right\rangle}\right]\left\langle \varepsilon \varphi\right\rangle
+2\frac{\delta\Gamma_2}{\delta \left\langle \psi^*\psi^* \right\rangle}\left\langle\psi^* \varphi\right\rangle\right\}
\label{45}
\te

\be
\left\langle \varphi \varphi^*\right\rangle=G\left\{\left[\omega^2\phi+2\frac{\delta\Gamma_2}{\delta \left\langle \psi^*\varepsilon\right\rangle}\right]\left\langle \varepsilon \varphi^*\right\rangle+2\frac{\delta\Gamma_2}{\delta \left\langle \psi^*\psi \right\rangle}\left\langle\psi \varphi^*\right\rangle\right\}
\label{46}
\te
where

\be
\left\langle \varepsilon \varphi^*\right\rangle=\left\langle \varepsilon \varphi\right\rangle^*=C\left[\omega^2\phi^*+2\left(\frac{\delta\Gamma_2}{\delta \left\langle \varepsilon \psi^*\right\rangle}\right)^*\right]G'^*
\label{47}
\te

\be
\left\langle\psi \varphi^*\right\rangle=-\left\langle\psi^* \varphi\right\rangle^*=-iG'^*
\label{48b}
\te
The reason for eq. (\ref{48b}) is that to obtain $\left\langle\psi \varphi^*\right\rangle$ from $\left\langle\psi^* \varphi\right\rangle$ we need both to take complex conjugate and change $\psi\to -\psi$ in the path integral.

\subsection{Charge conservation}
We wish to check if the conservation law eq. (\ref{17}) holds in the mean on shell. Observe that

\be
i\left\langle J^{\mu}\right\rangle\left(x\right)=\phi^*\left(x\right)\nabla^{\mu}\phi\left(x\right)
-\phi\left(x\right)\nabla^{\mu}\phi^*\left(x\right)+\left.\left[\nabla^{\mu}_x-\nabla^{\mu}_{x'}\right]\left\langle \varphi\left(x\right)\varphi^*\left(x'\right)\right\rangle\right|_{x'=x}
\label{54}
\te
Therefore, from eq. (\ref{25.2}),

\bea
&&i\nabla_{\mu}\left\langle J^{\mu}\right\rangle\left(x\right)+\phi^*j-\phi j^*=\nn
&&-\omega^2\phi^*\left\langle \varepsilon\varphi\right\rangle+\omega^2\phi\left\langle \varepsilon\varphi^*\right\rangle+\left[\mathbf{D}_0\left\langle \varphi\left(x\right)\varphi^*\left(x'\right)\right\rangle-\left\langle \varphi\left(x\right)\varphi^*\left(x'\right)\right\rangle\mathbf{D}_0\right]_{x'=x}
\label{56}
\tea
Now from the equations for the propagators we have

\bea
&&\omega^2\phi\left(x\right)\left\langle \varepsilon\left(x\right)\varphi^*\left(x'\right)\right\rangle+\mathbf{D}_0\left\langle \varphi\left(x\right)\varphi^*\left(x'\right)\right\rangle=\nn
&&-2\int\!dy\:\frac{\delta\Gamma_2}{\delta \left\langle \psi^*\left(x\right)X^s\left(y\right) \right\rangle}\left\langle X^s\left(y\right)\varphi^*\left(x'\right)\right\rangle\nn
&&\omega^2\left\langle \varphi\left(x\right)\varepsilon\left(x'\right)\right\rangle\phi^*\left(x'\right)+\left\langle \varphi\left(x\right)\varphi^*\left(x'\right)\right\rangle\mathbf{D}_0=\nn
&&-2\int\!dy\:\left\langle \varphi\left(x\right)X^s\left(y\right)\right\rangle\frac{\delta\Gamma_2}{\delta \left\langle X^s\left(y\right)\psi\left(x'\right) \right\rangle}
\label{58}
\tea
So conservation in the mean holds if

\be
\int\!dy\:\left\{\frac{\delta\Gamma_2}{\delta \left\langle \psi\left(x\right)X^s\left(y\right) \right\rangle}\left\langle \varphi\left(x\right)X^s\left(y\right)\right\rangle-\frac{\delta\Gamma_2}{\delta \left\langle \psi^*\left(x\right)X^s\left(y\right) \right\rangle}\left\langle\varphi^*\left(x\right) X^s\left(y\right)\right\rangle\right\}=0
\label{59}
\te
Eq. (\ref{59}) holds not only on shell, but identically. As we have seen, the 2PIEA is built from expectation values of powers of the expression

\be
\omega^2\int\!dx\:\varepsilon\left(\psi^*\varphi+\psi\varphi^*\right)
\label{60}
\te
which is obviously invariant under a transformation whereby 

\bea
\delta\psi&=&i\theta\varphi\nn
\delta\psi^*&=&-i\theta\varphi^*
\label{63}
\tea
Eq. (\ref{59}) is just the Zinn-Justin identity associated to this symmetry \cite{Zinn93,Cal04}, and therefore conservation holds.

\subsection{Intensity operator}
To investigate the structure of the 2PIEA it is crucial to observe that the Feynman graphs which actually contribute to it have a rather peculiar structure. Namely, any graph containing ``matter'' fields ($\varphi$, $\varphi^*$, $\psi$ or $\psi^*$) comes from taking the expectation value (in the Wick theorem sense) of some term in the power expansion of 

\be 
\left\langle e^{i\omega^2\int\!dx\:\varepsilon\left(\psi^*\varphi+\psi\varphi^*\right)}\right\rangle 
\label{60b}
\te
Let us look into any such terms, containing $N_{\varphi}$ $\varphi$ fields, and similarly for the $\psi$, $\varphi^*$ and $\psi^*$ fields. It is obvious that $N_{\varphi}=N_{\psi^*}$ and $N_{\varphi^*}=N_{\psi}$.

Now look at any Feynman graph obtained from the expectation value of that term. The graph will contain $L_{\left\langle \varphi\varphi\right\rangle } $ $\left\langle \varphi\varphi\right\rangle$ lines, and so on for all other (bosonic) propagators. The graph will be globally charge neutral, and also

\be 
N_{\varphi}=2L_{\left\langle \varphi\varphi\right\rangle} +L_{\left\langle \varphi\varphi^*\right\rangle} +L_{\left\langle \varphi\psi\right\rangle} +L_{\left\langle \varphi\psi^*\right\rangle} +L_{\left\langle \varphi\varepsilon\right\rangle }
\te
a similar analysis yields

\be 
N_{\psi^*}=L_{\left\langle \psi^*\varphi\right\rangle} +L_{\left\langle \psi^*\varphi^*\right\rangle} +L_{\left\langle \psi^*\psi\right\rangle} +2L_{\left\langle \psi^*\psi^*\right\rangle} +L_{\left\langle \psi^*\varepsilon\right\rangle }
\te
and so $N_{\varphi}=N_{\psi^*}$ implies

\be 
2L_{\left\langle \varphi\varphi\right\rangle} +L_{\left\langle \varphi\varphi^*\right\rangle} +L_{\left\langle \varphi\psi\right\rangle}  +L_{\left\langle \varphi\varepsilon\right\rangle }=L_{\left\langle \psi^*\varphi^*\right\rangle} +L_{\left\langle \psi^*\psi\right\rangle} +2L_{\left\langle \psi^*\psi^*\right\rangle} +L_{\left\langle \psi^*\varepsilon\right\rangle }
\te
Similarly

\be 
2L_{\left\langle \varphi^*\varphi^*\right\rangle} +L_{\left\langle \varphi\varphi^*\right\rangle} +L_{\left\langle \varphi^*\psi^*\right\rangle}  +L_{\left\langle \varphi^*\varepsilon\right\rangle }=L_{\left\langle \psi\varphi\right\rangle} +L_{\left\langle \psi\psi^*\right\rangle} +2L_{\left\langle \psi\psi\right\rangle} +L_{\left\langle \psi\varepsilon\right\rangle }
\te
To find $G=i\left\langle \varphi\psi^*\right\rangle$ and $G^*$ we must solve eq. (\ref{40c})
and its conjugate. Now any graph contributing to the variational derivative must have 

\be
L_{\left\langle \psi\psi\right\rangle }=L_{\left\langle \psi\psi^*\right\rangle }=L_{\left\langle \psi^*\psi^*\right\rangle }=L_{\left\langle \psi^*\varepsilon\right\rangle}=L_{\left\langle \psi\varepsilon\right\rangle}=L_{\left\langle \psi^*\varphi^*\right\rangle}=L_{\left\langle \psi\varphi\right\rangle}=0
\label{25.7c}
\te 
because the presence of any of those propagators would kill the graph on shell. It follows that it also must have

\be 
L_{\left\langle \varphi\varphi\right\rangle} =L_{\left\langle \varphi\varphi^*\right\rangle} =L_{\left\langle \varphi\psi\right\rangle}  =L_{\left\langle \varphi\varepsilon\right\rangle }=L_{\left\langle \varphi^*\varphi^*\right\rangle} =L_{\left\langle \varphi\varphi^*\right\rangle} =L_{\left\langle \varphi^*\psi^*\right\rangle}  =L_{\left\langle \varphi^*\varepsilon\right\rangle }=0
\te 
The only propagators left to build the graph with are $G$, $G^*$ and $C$, and so we obtain a closed dynamics.

Observe that because of charge conservation at every vertex, matter fields (defined above) can only appear in closed matter loops, connected by $C$ lines to themselves, other mater loops or ghost loops. Actually graphs containing more than one matter loop cancel out in the derivative against the contribution of the graphs where each matter loop is replaced in turn by a ghost loop (a ghost loop having the same amplitude than a matter loop but inverse sign), so that only graphs with a single matter loop need to be considered. We shall discuss graphs containing $C$ loops below.

Let us turn to $\left\langle \varphi\left(x\right)\varepsilon\left(x'\right)\right\rangle$ and $\left\langle \varphi^*\left(x\right)\varepsilon\left(x'\right)\right\rangle$. Now we must solve Eq. (\ref{43}), so must seek graphs with $L_{\left\langle \varepsilon \psi^*\right\rangle}=1$. Therefore we shall have 

\be 
2L_{\left\langle \varphi\varphi\right\rangle} +L_{\left\langle \varphi\varphi^*\right\rangle} +L_{\left\langle \varphi\psi\right\rangle}  +L_{\left\langle \varphi\varepsilon\right\rangle }=1
\te 
Now, because $\left\langle \varepsilon \psi^*\right\rangle$ has charge $-1$, the remainder of the graph containing it must have charge $+1$. Therefore we must have 

\be 
L_{\left\langle \varphi\varepsilon\right\rangle }=1
\te 
and all other zero

\be 
L_{\left\langle \varphi\varphi\right\rangle} =L_{\left\langle \varphi\varphi^*\right\rangle} =L_{\left\langle \varphi\psi\right\rangle} =0
\te 
It follows that again we get a closed equation for $\left\langle \varepsilon\varphi\right\rangle$, and moreover this equation is linear. 

A similar analysis shows that we get a linear, self-consistent equation for $\left\langle \varphi\varphi^*\right\rangle $. The equation to be solved is (\ref{46}). 
Since we have already analyzed graphs with $L_{\left\langle \varepsilon \psi^*\right\rangle}=1$, let us focus on the third term. We seek Feynman graphs with $L_{\left\langle \psi^*\psi\right\rangle}=1$, since any such propagator surviving the derivative will kill the graph. This leaves two options, either $L_{\left\langle \varphi\varphi^*\right\rangle}=1$ and $L_{\left\langle \varphi\varepsilon\right\rangle }=L_{\left\langle \varphi^*\varepsilon\right\rangle }=0$ or else $L_{\left\langle \varphi\varphi^*\right\rangle}=0$ and $L_{\left\langle \varphi\varepsilon\right\rangle }=L_{\left\langle \varphi^*\varepsilon\right\rangle }=1$. Therefore the third term may be written as the sum of two terms, one containing only $C$, $G$, $G^*$, $\left\langle \varepsilon\varphi\right\rangle $ and $\left\langle \varepsilon\varphi^*\right\rangle $, and the other linear in $\left\langle \varphi\varphi^*\right\rangle $. Observe furthermore that $\left\langle \varphi^*\psi\right\rangle =-iG^*$ (recall eq. (\ref{48b})). If we then compare eq. (\ref{46}) with eq. (\ref{16}), we conclude that we may identify the intensity operator with the operator in the term in eq. (\ref{46}) which is linear in $\left\langle \varphi\varphi^*\right\rangle $, namely

\bea 
&&I\left[z,y;z',y'\right]=\nn
&&-2i\left\lbrace \frac{\delta^2\Gamma_2}{\delta \left\langle \psi^*\left( z\right) \psi\left( z'\right)   \right\rangle\delta\left\langle \varphi\left( y\right)\varphi^*\left( y'\right)  \right\rangle}+\frac{\delta^2\Gamma_2}{\delta \left\langle \psi^*\left( z\right) \psi\left( z'\right)   \right\rangle\delta\left\langle \varphi^*\left( y'\right)\varphi\left( y\right)  \right\rangle}\right\rbrace 
\label{intop}
\tea
The two derivatives are necessary because in building the 2PI EA $\left\langle \varphi\varphi^*\right\rangle $ and $\left\langle \varphi^*\varphi\right\rangle $ must be regarded as independent variables, although they will be equal to each other ``on shell'', namely, when evaluated on a physical solution.

\section{Loop expansion of the 2PI EA}
In this section we shall discuss in some detail the first three orders ($L=1$, $2$ and $3$) in the loop expansion of the 2PI EA. To this effect, it is convenient to introduce a single ``matter'' field multiplet $\chi^a=\left(\varphi ,\varphi^*,\psi ,\psi^* \right) $ and a ``ghost'' field multiplet $\gamma^a=\left(\eta ,\vartheta,\zeta ,\xi \right) $. In terms of these fields the quantum part of the effective action $\Gamma_2$ may be written as

\be 
e^{i\Gamma_2}=\left\langle e^{\frac{i}{2}\omega^2\int\!dx\:\varepsilon\sigma_{ab}\left(\chi^a\chi^b+i\gamma^a\gamma^b\right)-\sum_{n=3}^{\infty}\frac{1}{n!}\int\;dx_1...dx_n\;f_n\left( x_1,...,x_n\right) \varepsilon\left( x_1\right)... \varepsilon\left( x_n\right) }\right\rangle _{2PI}
\label{48}
\te 
where

\be 
\sigma_{ab}=\delta^1_a\delta^4_b+\delta^2_a\delta^3_b+\delta^3_a\delta^2_b+\delta^4_a\delta^1_b
\te 
Variations of the effective action are derived from the  rule

\be 
\frac{\delta\left\langle \chi^a\left( z_1\right) \chi^b\left( z_2\right)\right\rangle }{\delta\left\langle \chi^c\left( x_1\right) \chi^d\left( x_2\right)\right\rangle}= \delta^a_c\delta\left( z_1-x_1\right)\delta^b_d\delta\left( z_2-x_2\right)
\label{rule}
\te
The $2PI$ subscript in eq. (\ref{48}) means that to compute the 2PI EA the exponential is expanded into its power series, then each expectation value is computed according to Wick's theorem, using full propagators in the internal lines, and finally non 2PI graphs are discarded.

The first observation is that the graphs so built have no external lines. Let us introduce the numbers  $V_n$ of $n$-point vertices, $I$ of lines and $L$ of loops.

\bea 
\sum nV_n&=&2I\nn
I-V&=&L-1
\tea
Elliminating $I$ between these equations we obtain

\be 
\sum \left( n-2\right) V_n=2\left( L-1\right) 
\te 
Therefore, if we consider graphs up to a certain number $L$ of loops, only vertices with $n\le 2L$ need to be considered, and only those with $n\le L+1$ will appear more than once in the same graph \cite{CalHu95a}. This fact makes the loop expansion the simplest perturbative approach to the evaluation of the 2PIEA.

The lowest (one loop) approximation corresponds to $L=1$. Therefore $V_n=0$ for all $n$ and  $\Gamma^{\left( 1\right) }_2=0$. From eq. (\ref{41}) $\Sigma=Q=0$. This is just the ``classical'' approximation where only coherent fields are considered.

\subsection{Two loops theory}

The following approximation is the two-loops one, where $L=2$. This means $V_3=2$ and $V_n=0$ for $n>3$, or else $V_3=0$, $V_4=1$ and $V_n=0$ for $n>4$. Therefore we may write

\bea
\Gamma^{\left( 2\right) } _2&=&\frac {-i}2\left\langle \left\{\frac i2\omega^2\int\!dx\:\varepsilon\sigma_{ab}\left(\chi^a\chi^b+i\gamma^a\gamma^b\right)\right.\right.\nn
 &-&\left.\left.\frac{1}{6}\int\;dx_1dx_2dx_3\;f_3\left( x_1,x_2,x_3\right) \varepsilon\left( x_1\right)\varepsilon\left( x_2\right) \varepsilon\left( x_3\right)\right\}^2 \right\rangle _{2PI}\nn
&+&\frac{i}{24}\left\langle \int\;dx_1dx_2dx_3dx_4\;f_4\left( x_1,x_2,x_3,x_4\right) \varepsilon\left( x_1\right)\varepsilon\left( x_2\right) \varepsilon\left( x_3\right) \varepsilon\left( x_4\right)\right\rangle _{2PI}
\label{49}
\tea
To compute the self energy and the intensity operator we only require the Feynman graphs carrying matter fields in their internal lines. These are

\be 
\Gamma^{\left( 2\right) } _{2,matter}=\frac i4\omega^4\int\;dz_1dz_2\;C\left( z_1,z_2\right) \sigma_{a_1b_1}\sigma_{a_2b_2}\left\langle \chi^{a_1}\left( z_1\right) \chi^{a_2}\left( z_2\right)\right\rangle 
\left\langle \chi^{b_1}\left( z_1\right) \chi^{b_2}\left( z_2\right)\right\rangle 
\te 
We now derive the self energy from eq. (\ref{41}) and the intensity operator from eq. (\ref{intop}). Computing the variation of the EA with the rule eq. (\ref{rule}) we immediately find the nonlinear and ladder approximations, respectively. We see that to this order there are no contributions from the nongaussianity in the $\varepsilon$ fluctuations; these will show up at the next order.

\subsection{Three loops theory}
At three loops $L=3$ the nontrivial possibilities are $V_3=4$, all others zero, $V_3=2$, $V_4=1$, all others zero, or else $V_4=2$, all others zero. Therefore

\bea
\Gamma^{\left( 3\right) } _2&=&\frac {-i}{24}\left\langle \left\{\frac i2\omega^2\int\!dx\:\varepsilon\sigma_{ab}\left(\chi^a\chi^b+i\gamma^a\gamma^b\right)\right.\right.\nn
 &-&\left.\left.\frac{1}{6}\int\;dx_1dx_2dx_3\;f_3\left( x_1,x_2,x_3\right) \varepsilon\left( x_1\right)\varepsilon\left( x_2\right) \varepsilon\left( x_3\right)\right\}^4 \right\rangle _{2PI}\nn
 &+& \frac {i}{2}\left\langle \left\{\frac i2\omega^2\int\!dx\:\varepsilon\sigma_{ab}\left(\chi^a\chi^b+i\gamma^a\gamma^b\right)\right.\right.\nn
 &-&\left.\frac{1}{6}\int\;dx_1dx_2dx_3\;f_3\left( x_1,x_2,x_3\right) \varepsilon\left( x_1\right)\varepsilon\left( x_2\right) \varepsilon\left( x_3\right)\right\}^2 \nn
 &&\left. \int\;dy_1dy_2dy_3dy_4\;f_4\left( y_1,y_2,y_3,y_4\right) \varepsilon\left( y_1\right)\varepsilon\left(y_2\right) \varepsilon\left(y_3\right) \varepsilon\left( y_4\right)\right\rangle _{2PI}\nn
&-&\frac{i}{2}\left\langle \left\lbrace \int\;dx_1dx_2dx_3dx_4\;f_4\left( x_1,x_2,x_3,x_4\right) \varepsilon\left( x_1\right)\varepsilon\left( x_2\right) \varepsilon\left( x_3\right) \varepsilon\left( x_4\right)\right\rbrace ^2\right\rangle _{2PI}
\label{49b}
\tea
The terms that contribute to the self energy and the intensity operator can be written as

\bea 
&&\Gamma^{\left( 3\right) } _{2,matter}=A+B\nn
&&A=\frac {-i}{24} \frac{\omega^8}{16}\left\langle \left[\int\!dx\:\varepsilon\sigma_{ab}\chi^a\chi^b \right] ^4\right\rangle _{2PI} \nn
&&B=\frac {1}{24}  \frac{\omega^6}{12}\left\langle \left[\int\!dx\:\varepsilon\sigma_{ab}\chi^a\chi^b \right] ^3\int\;dx_1dx_2dx_3\;f_3\left( x_1,x_2,x_3\right) \varepsilon\left( x_1\right)\varepsilon\left( x_2\right) \varepsilon\left( x_3\right)\right\rangle _{2PI}\nn
\tea 
Explicitly

\bea
&&A=\frac{-i\omega^8}{8}\int\;dx_1dx_2dx_3dx_4\;\sigma_{a_1b_1}\sigma_{a_2b_2}\sigma_{a_3b_3}\sigma_{a_4b_4}C\left(x_1,x_4\right)C\left(x_2,x_3\right)\nn
&&\left\langle \chi^{a_1}\left(x_1\right)\chi^{a_2}\left(x_2\right)\right\rangle\left\langle \chi^{b_1}\left(x_1\right)\chi^{a_3}\left(x_3\right)\right\rangle\left\langle \chi^{b_2}\left(x_2\right)\chi^{a_4}\left(x_4\right)\right\rangle\left\langle \chi^{b_3}\left(x_3\right)\chi^{b_4}\left(x_4\right)\right\rangle\nn
&&B=\frac{-\omega^6}{6}\int\;dx_1dx_2dx_3\;\sigma_{a_1b_1}\sigma_{a_2b_2}\sigma_{a_3b_3}K\left(x_1,x_2,x_3\right)\nn
&&\left\langle \chi^{a_1}\left(x_1\right)\chi^{a_2}\left(x_2\right)\right\rangle\left\langle \chi^{b_1}\left(x_1\right)\chi^{a_3}\left(x_3\right)\right\rangle\left\langle \chi^{b_2}\left(x_2\right) \chi^{b_3}\left(x_3\right)\right\rangle\nn
&&K\left(x_1,x_2,x_3\right)=-\int\;dy_1dy_2dy_3\;f_3\left( y_1,y_2,y_3\right)C\left(x_1,y_1\right)C\left(x_2,y_2\right)C\left(x_3,y_3\right)
\tea
To compute the $K$ kernel we observe that

\be
\frac{\delta\Gamma_2}{\delta f_3\left( x_1,x_2,x_3\right) }=\frac i6\left\langle \varepsilon\left( x_1\right)\varepsilon\left( x_2\right) \varepsilon\left( x_3\right)\right\rangle_{2PI}
\te
To leading order we may substitute the 2-loops expression eq. (\ref{49}) for the effective action, whereby we find

\be 
K\left(x_1,x_2,x_3\right)=\left\langle \varepsilon\left( x_1\right)\varepsilon\left( x_2\right) \varepsilon\left( x_3\right)\right\rangle_{2PI}
\te
therefore $K$ would vanish in a gaussian theory. The self energy and intensity operator have a similar structure

\be
\Sigma\left( x_1,x_4\right)    = \Sigma_A\left( x_1,x_4\right) +\Sigma_B\left( x_1,x_4\right) 
\te 

\be 
\Sigma_A\left( x_1,x_4\right) =\omega^8\int\;dx_2dx_3\;C\left(x_1,x_3\right)C\left(x_2,x_4\right)G\left(x_1,x_2\right)G\left(x_2,x_3\right)
G\left(x_3,x_4\right)
\label{sigma2g}
\te 

\be 
\Sigma_B\left( x_1,x_4\right) = \omega^6\int\;dx_2\;K\left(x_1,x_2,x_4\right)G\left(x_1,x_2\right)G\left(x_2,x_4\right)
\label{sigma2ng}
\te

\bea
I\left[z,y;z',y'\right]&=&I_A\left[z,y;z',y'\right]+I_B\left[z,y;z',y'\right]\nn
I_A\left[z,y;z',y'\right]&=&I_A^{\left( 1\right) }\left[z,y;z',y'\right]+I_A^{\left( 2\right)}\left[z,y;z',y'\right]+I_A^{\left( 3\right) }\left[z,y;z',y'\right] \nn
I_B\left[z,y;z',y'\right]&=&I_B^{\left( 1\right) }\left[z,y;z',y'\right]+I_B^{\left( 2\right)}\left[z,y;z',y'\right]\nn
I_A^{\left( 1\right) }\left[z,y;z',y'\right]&=&\omega^8\delta\left( z-y\right) C\left( z',y'\right) \int\;dx\;C\left( z,x\right) G^*\left( z',x\right) G^*\left( x,y'\right) \nn
I_A^{\left( 2\right) }\left[z,y;z',y'\right]&=&\omega^8\delta\left( z'-y'\right) C\left( z,y\right) \int\;dx\;C\left( x,z'\right) G\left( z,x\right) G\left( x,y\right) \nn
I_A^{\left( 3\right) }\left[z,y;z',y'\right]&=&\omega^8C\left( z,y'\right)C\left( z',y\right)G\left( z,y\right)G^*\left( z',y'\right)\nn
I_B^{\left( 1\right) }\left[z,y;z',y'\right]&=&\omega^6\delta\left( z-y\right) K\left(z,z',y'\right) G^*\left( z',y'\right) \nn
I_B^{\left( 2\right) }\left[z,y;z',y'\right]&=&\omega^6\delta\left( z'-y'\right) K\left(z,z',y\right) G\left( z,y\right)
\label{2lint}
\tea

\section{Application: overlapping intrusions on a homogeneous background}

As an application, we will consider the case of a homogeneous medium with $\epsilon=\epsilon_0$, on which there are introduced spherical, overlapping bubbles with $\epsilon=\epsilon_1=\epsilon_0+\Delta\epsilon$. The bubbles all have the same radius $R$, and their centers are chosen independently at random with a homogeneous distribution over some control volume $V$. Let there be $N$ bubbles centered at points $\left( \mathbf{x}_1,...\mathbf{x}_N\right) $. Then 

\be
\epsilon\left( \mathbf{x}\right) =\epsilon_0-\Delta \epsilon \sum_{k=1}^{N}{\left( -1\right) ^k}\sum_{i_1< i_2<...<i_k}f_{i_1}\left( \mathbf{x}\right)...f_{i_k}\left( \mathbf{x}\right)
\te
where

\be 
f_i\left( \mathbf{x}\right)=\theta\left( 1-\frac{\left(  \mathbf{x}- \mathbf{x}_i\right) ^2}{R^2}\right)\equiv f\left(  \mathbf{x}- \mathbf{x}_i\right) 
\te 
Given this representation, we easily find

\be 
\bar{\epsilon}=\epsilon_0+\Delta \epsilon \left(1- \left( 1- \frac vV\right)^N\right) 
\te
where $v=4\pi R^3/3$ is the volume of one bubble. We define 

\be 
\varepsilon\left( \mathbf{x}\right) =\epsilon\left( \mathbf{x}\right) -\bar{\epsilon}=\epsilon\left( \mathbf{x}\right) -\Delta \epsilon+\Delta \epsilon\left( 1- \frac vV\right)^N
\te
Then

\be
C\left( \mathbf{x},\mathbf{x}'\right) = -\varepsilon_0^2+\Delta \epsilon^2 \left( 1+ \frac vV\left( \gamma\left( \mathbf{x},\mathbf{x}'\right)-2\right) \right) ^{N}
\te
where 

\be 
\varepsilon_0=-\Delta \epsilon  \left( 1- \frac vV\right)^N
\te
and we have defined

\be 
\left\langle f_{1}\left( \mathbf{x}\right)f_{1}\left(\mathbf{x}'\right)\right\rangle =\frac vV \gamma\left( \mathbf{x},\mathbf{x}'\right)
\te 
$v\gamma$ is the volume of the intersection between an sphere centered at $\mathbf{x}$ and another centered at $\mathbf{x}'$. It is clear that $\gamma\left( 0\right) =1$ and $\gamma\left( \mathbf{x}\right)=0$ if $\left(\mathbf{x-x'}^2\right)\ge 4R^2$. 

For the three point correlations we get

\bea 
K\left( \mathbf{x},\mathbf{x}',\mathbf{x}''\right)& =&-\Delta \epsilon^3 \left( 1-\frac vV\left(\lambda\left( \mathbf{x},\mathbf{x}',\mathbf{x}''\right)-\gamma\left( \mathbf{x},\mathbf{x}'\right) -\gamma\left( \mathbf{x},\mathbf{x}''\right) -\gamma\left( \mathbf{x}',\mathbf{x}''\right) +3\right) \right) ^N\nn
& -&\varepsilon_0\left[C\left( \mathbf{x},\mathbf{x}'\right)+C\left( \mathbf{x},\mathbf{x}''\right)+C\left( \mathbf{x}',\mathbf{x}''\right)\right]  -\varepsilon_0^3
\tea 
where we defined

\be 
\left\langle f_{1}\left( \mathbf{x}\right)f_{1}\left( \mathbf{x}'\right)f_{1}\left( \mathbf{x}''\right)\right\rangle =\frac vV\lambda\left( \mathbf{x},\mathbf{x}',\mathbf{x}''\right)
\te
$v\lambda$ is the volume of the intersection of three spheres. The fact that $K$ is non vanishing shows that $\varepsilon$ fluctuations are not Gaussian.

We are interested in the limit where $N,V\to\infty$, $N/V=\rho=$ constant. In this limit

\be 
\bar{\epsilon}=\epsilon_0+\Delta \epsilon \left(1- e^{-\rho v}\right) 
\te

\be
C\left( \mathbf{x},\mathbf{x}'\right) =\Delta \epsilon^2e^{-2\rho v}  \left[e^{\rho v \gamma\left( \mathbf{x},\mathbf{x}'\right)}-1\right]
\te
$C$ vanishes when the distance between its arguments exceeds $2R$, and

\bea 
K\left( \mathbf{x},\mathbf{x}',\mathbf{x}''\right)& =&-\Delta \epsilon^3e^{-3\rho v} \left\{e^{-\rho v\left(\lambda\left( \mathbf{x},\mathbf{x}',\mathbf{x}''\right)-\gamma\left( \mathbf{x},\mathbf{x}'\right) -\gamma\left( \mathbf{x},\mathbf{x}''\right) -\gamma\left( \mathbf{x}',\mathbf{x}''\right) \right) }\right.\nn
& -&\left.\left[e^{\rho v \gamma\left( \mathbf{x},\mathbf{x}'\right)}+e^{\rho v \gamma\left( \mathbf{x},\mathbf{x}''\right)}+e^{\rho v \gamma\left( \mathbf{x}',\mathbf{x}''\right)}\right] +2\right\}
\tea 
$K$ vanishes when the distance between any two arguments exceeds $2R$. $\rho v$ is the fraction of the total volume occupied by the spheres, not counting  overlapping. If they are sparse, then $\rho v\ll 1$, In this limit, the formulae above simplify further to

\be 
\bar{\epsilon}=\epsilon_0+\Delta \epsilon \rho v 
\te

\be
C\left( \mathbf{x},\mathbf{x}'\right) =\Delta \epsilon^2\rho v \gamma\left( \mathbf{x},\mathbf{x}'\right)
\label{sparseC}
\te

\be
K\left( \mathbf{x},\mathbf{x}',\mathbf{x}''\right)=\Delta \epsilon^3\rho v\lambda\left( \mathbf{x},\mathbf{x}',\mathbf{x}''\right)
\label{sparseK}
\te 
\subsection{Non-gaussian contributions to the diffuse intensity}
We shall conclude by analyzing in some detail the non-gaussian contributions to the diffuse intensity. This is measured by the two-point correlation

\be 
G_1\left( \mathbf{x},\mathbf{x}'\right) =\left\langle \varphi\left(\mathbf{x}\right) \varphi^*\left( \mathbf{x}'\right) \right\rangle 
\te 
If we assume the mean field is just a plane wave $\phi=\left( 2\pi\right) ^{-3/2}\phi_0 e^{i\mathbf{K}\mathbf{x}}$, then the problem is translation invariant and we may Fourier transform all relevant quantities, e.g.

\be 
G_1\left( x,x'\right) =\int\;\frac{d^3p}{\left( 2\pi\right) ^3}\;e^{i\mathbf{p}\left( \mathbf{x}-\mathbf{x}'\right) }G_1\left( \mathbf{p}\right) 
\te 
We similarly introduce the Fourier transforms $G\left( \mathbf{p}\right)$ and $Q\left( \mathbf{p}\right)$ for the propagator and the self energy, respectively. The intensity operator becomes

\be 
I\left[ \mathbf{z},\mathbf{y}; \mathbf{z}',\mathbf{y}'\right] =\int\;\frac{d^3p}{\left( 2\pi\right) ^3}\frac{d^3p'}{\left( 2\pi\right) ^3}\frac{d^3q}{\left( 2\pi\right) ^3}\;e^{i\mathbf{p}\left( \mathbf{z}-\mathbf{y}\right) }e^{-i\mathbf{p}'\left( \mathbf{z}'-\mathbf{y}'\right) }e^{i\mathbf{q}\left( \mathbf{y}-\mathbf{y}'\right) }I\left[ \mathbf{p},\mathbf{p}'; \mathbf{q}\right]
\te 
The Bethe-Salpeter equation reads

\be 
G_1\left( \mathbf{p}\right) =\mid G\left( \mathbf{p}\right)\mid^2\left\lbrace \phi_0^2\mathcal{I}\left[ \mathbf{p},\mathbf{K}\right]+\int\;\frac{d^3q}{\left( 2\pi\right) ^3}\mathcal{I}\left[ \mathbf{p},\mathbf{q}\right]G_1\left( \mathbf{q}\right)\right\rbrace 
\label{BeSal}
\te
where 

\be 
\mathcal{I}\left[ \mathbf{p},\mathbf{K}\right]={I}\left[ \mathbf{p},\mathbf{p}; \mathbf{p-K}\right]
\te
Since flux is conserved, for there to be scattered intensity the mean field equation must display absorption, namely, $Q\left( \mathbf{p}\right)$ must be complex. Indeed, the Ward identity reads

\be 
\mathrm{Im}Q\left( \mathbf{q}\right)=\int\;\frac{d^3p}{\left( 2\pi\right) ^3}\;\mathrm{Im}G\left( \mathbf{p}\right)\mathcal{I}\left[ \mathbf{p},\mathbf{q}\right]
\te 
where, moreover, since $G\left( \mathbf{p}\right)=\left[ p^2-\bar{\epsilon}\omega^2-Q\left( \mathbf{p}\right)\right] ^{-1}$,

\be 
\mathrm{Im}G\left( \mathbf{p}\right)=\mid G\left( \mathbf{p}\right)\mid^2\mathrm{Im}Q\left( \mathbf{p}\right)
\te 
Each term in the perturbative expansion of $I$ (eqs. (\ref{ladder}) and (\ref{2lint})) yields a corresponding term in the expansion of $\mathcal{I}$. Let us expand on the structure of some of these terms.

The leading Gaussian contribution to the scattered intensity is given by the ladder approximation

\be 
I\left[ \mathbf{p},\mathbf{p}'; \mathbf{q}\right]=\omega^4C\left(\mathbf{q}\right) 
\te 
Observe that

\be 
v\gamma\left( \mathbf{x},\mathbf{x}'\right)=\left( 2\pi\right) ^3\int\;\frac{d^3p}{\left( 2\pi\right) ^3}\frac{d^3p'}{\left( 2\pi\right) ^3}\;e^{i\left( \mathbf{p} \mathbf{x}+\mathbf{p}' \mathbf{x}'\right)  } f\left(  \mathbf{p}\right) f\left( \mathbf{p}'\right) \delta\left( \mathbf{p}+\mathbf{p}'\right) 
\te 
where 

\be 
f\left(  \mathbf{p}\right) =\frac{3v}{\left( Rp\right)^3 }\left[ \sin\left( Rp\right) -Rp\cos\left( Rp\right) \right] 
\te
$f\left( \mathbf{p}\right)$ is esentially flat (and equal to $v$) up to wave numbers of order $1/R$. Then the approximation eq. (\ref{sparseC}) gives

\be 
I_{ladder}\left[ \mathbf{p},\mathbf{p}'; \mathbf{q}\right]=\omega^4\Delta \epsilon^2\rho  f\left( \mathbf{q}\right)^2
\te  

\begin{figure}[htb]
\begin{center}
\includegraphics[scale=0.46]{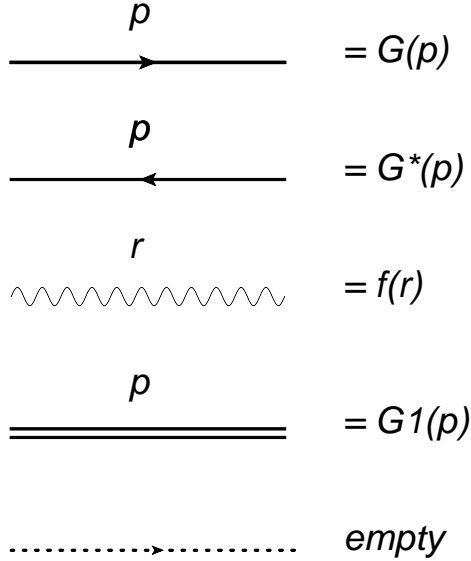}
\caption{The graphic code we shall use in this paper.  We represent $G$ by a right pointing arrow, $G^*$ by a left pointing one, $f$ by a wavy line and $G_1$ by a double arrow. Points indicate that no propagator is assigned to that line. }
\label{code}
\end{center}
\end{figure}

\begin{figure}[htb]
\begin{center}
\includegraphics[scale=0.6]{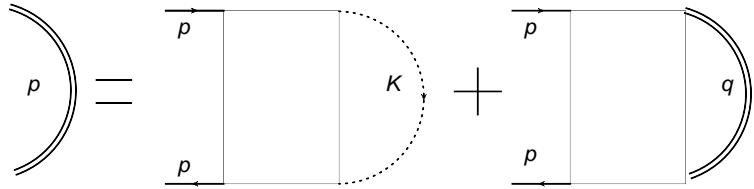}
\caption{The Bethe-Salpeter equation (\ref{BeSal}) in graphic representation }
\label{besalg}
\end{center}
\end{figure}

\begin{figure}[htb]
\begin{center}
\includegraphics[scale=0.6]{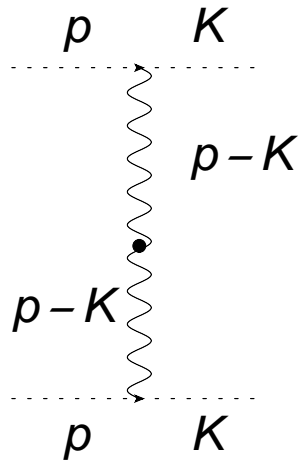}
\caption{The term in $\mathcal{I}$ from the ladder approximation  }
\label{ladderg}
\end{center}
\end{figure}

\begin{figure}[htb]
\begin{center}
\includegraphics[scale=0.6]{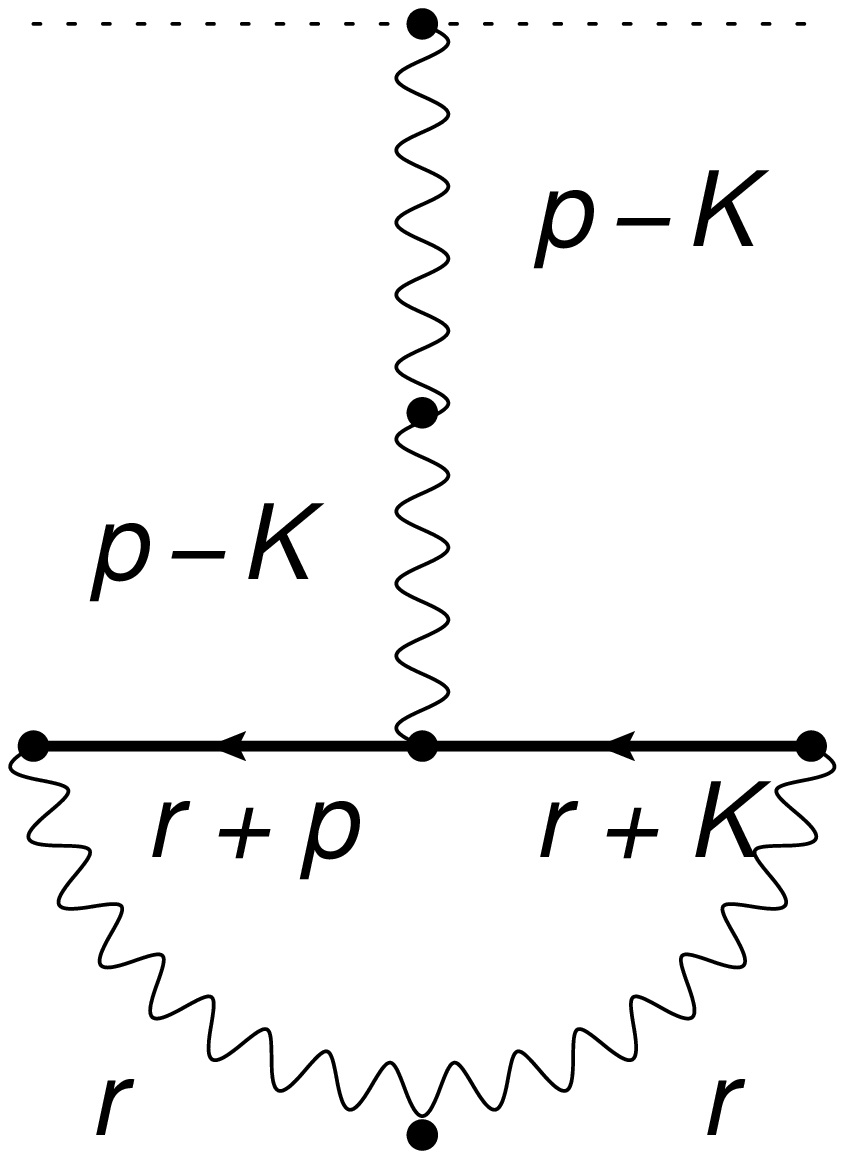}
\caption{The term in $\mathcal{I}$ from $I_A^{\left(1\right)} $ in eq. (\ref{2lint}). }
\label{a1g}
\end{center}
\end{figure}

\begin{figure}[htb]
\begin{center}
\includegraphics[scale=0.6]{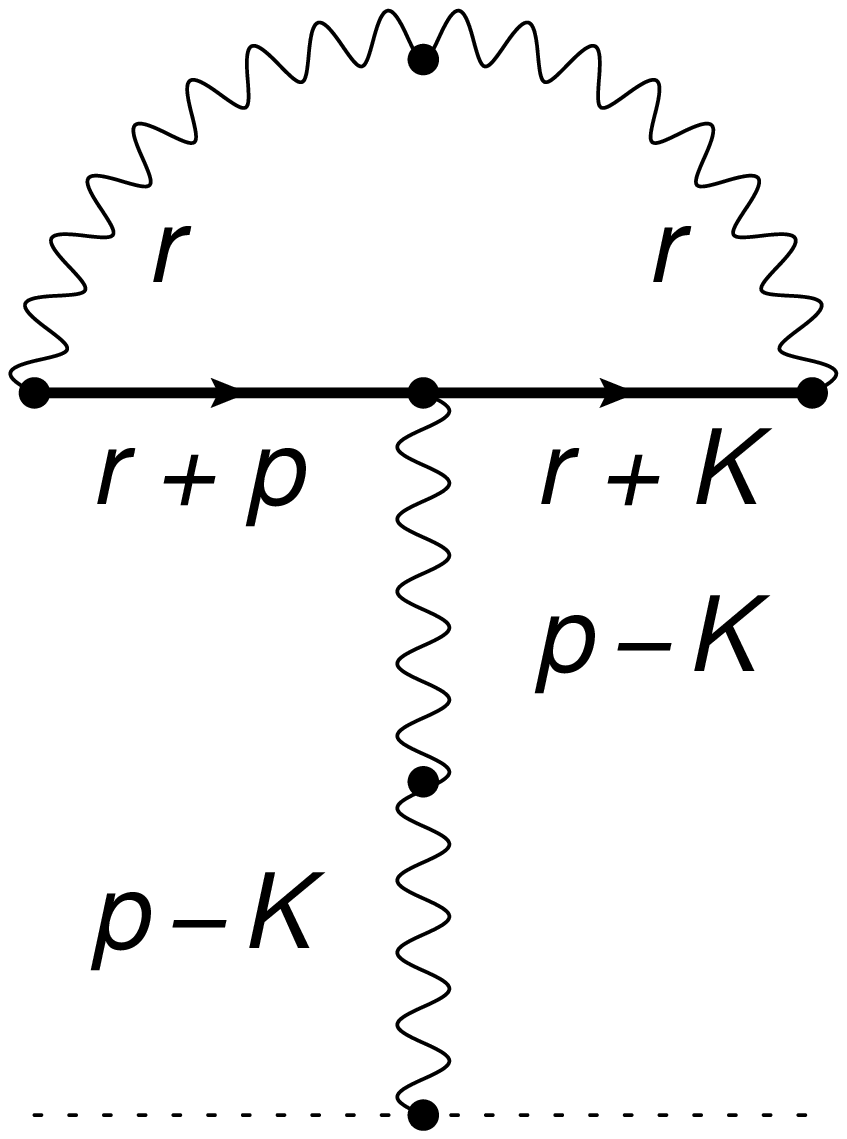}
\caption{The term in $\mathcal{I}$ from $I_A^{\left(2\right)} $ in eq. (\ref{2lint}). }
\label{a2g}
\end{center}
\end{figure}

\begin{figure}[htb]
\begin{center}
\includegraphics[scale=0.6]{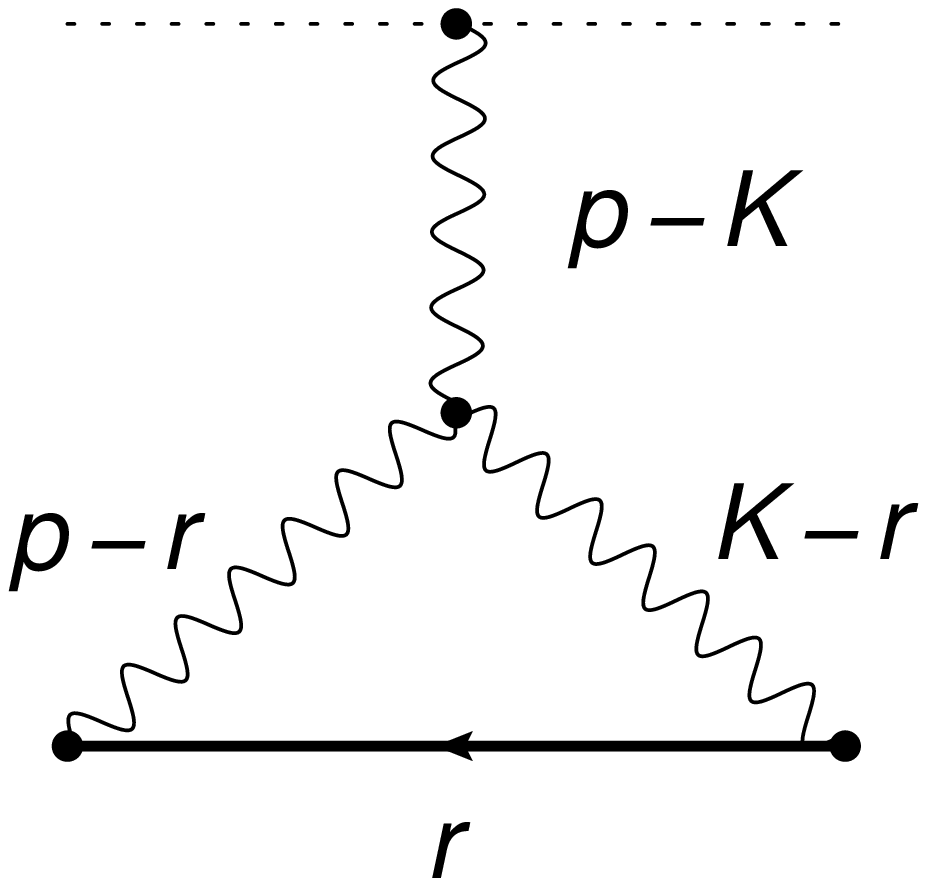}
\caption{The non-gaussian term in $\mathcal{I}$ from $I_B^{\left(1\right)} $ in eq. (\ref{2lint}). }
\label{blg}
\end{center}
\end{figure}

\begin{figure}[htb]
\begin{center}
\includegraphics[scale=0.6]{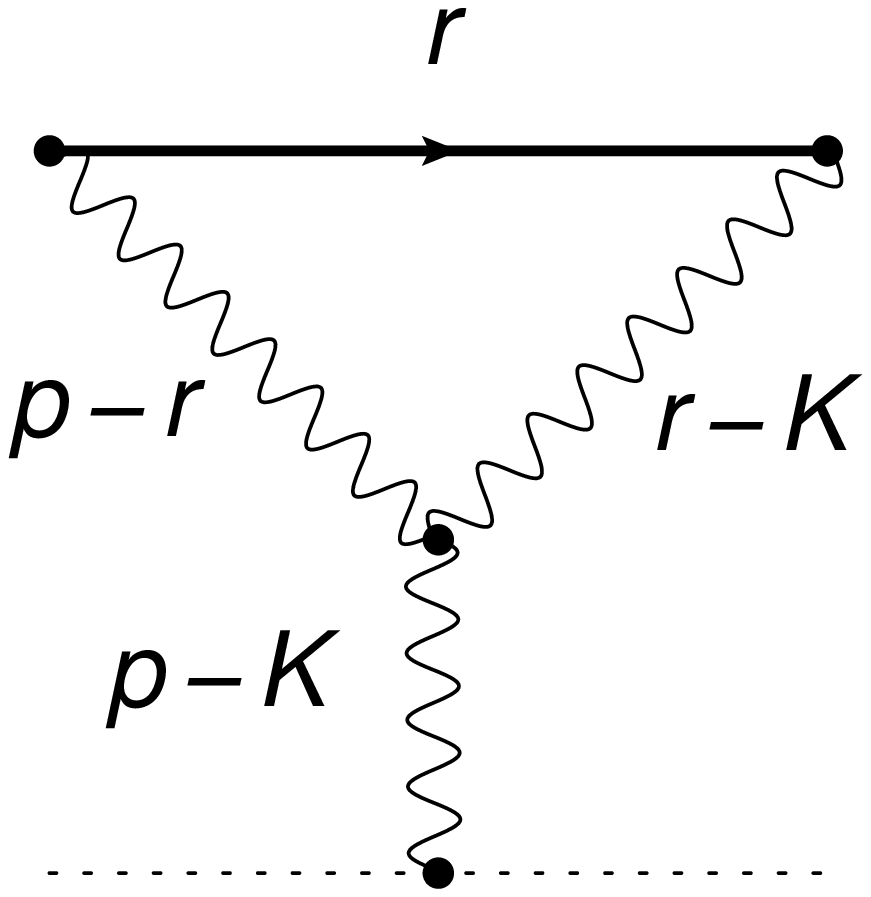}
\caption{The non-gaussian term in $\mathcal{I}$ from $I_B^{\left(2\right)} $ in eq. (\ref{2lint}). }
\label{b2g}
\end{center}
\end{figure}

\begin{figure}[htb]
\begin{center}
\includegraphics[scale=0.6]{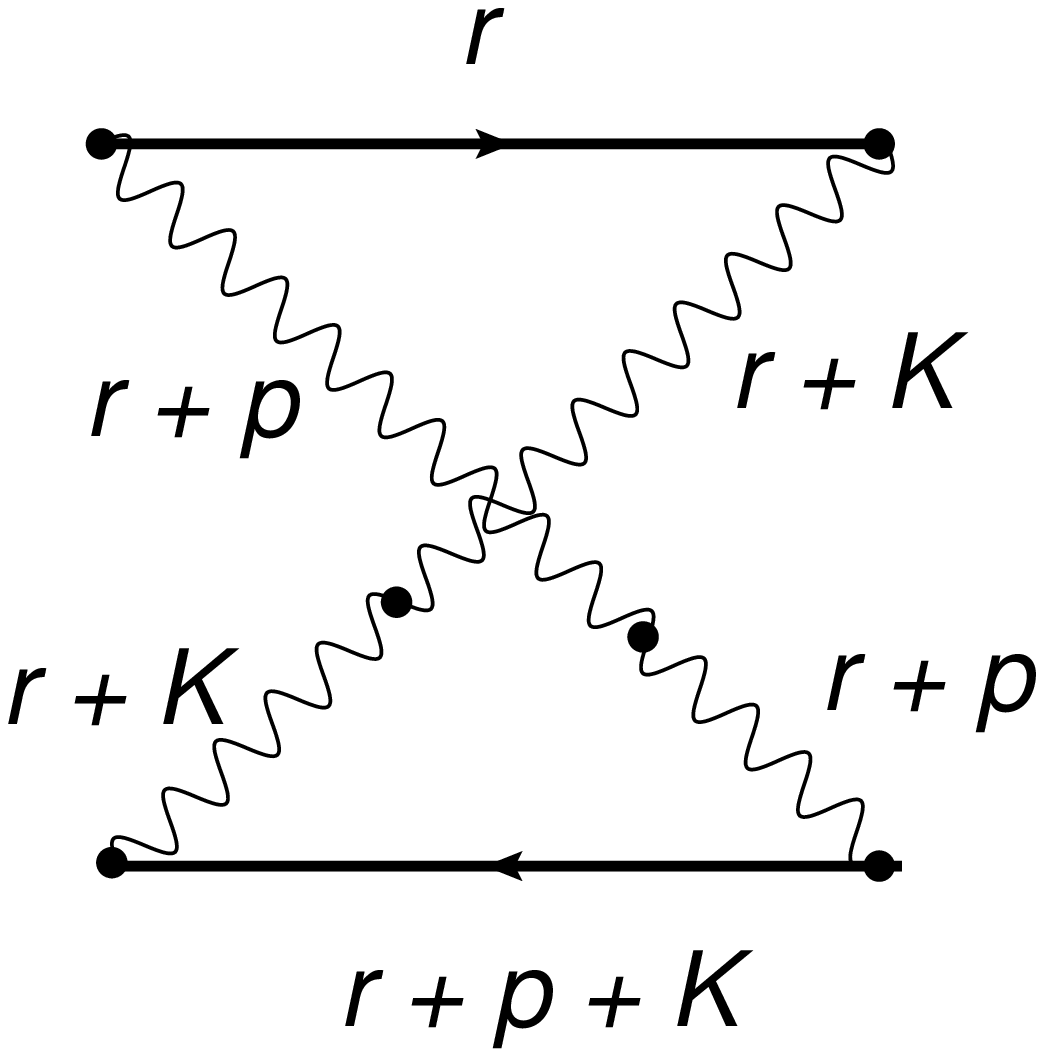}
\caption{The term in $\mathcal{I}$ from the maximally crossed graph $I_A^{\left(3\right)} $ in eq. (\ref{2lint}). Observe that $\mathbf{K}$ and $\mathbf{p}$ appear as $\mathbf{K+p}$; no other graph has this feature at this order in the loop expansion. }
\label{a3g}
\end{center}
\end{figure}

The nongaussian contributions to the intensity operator are given by the last two lines in eq. (\ref{2lint}). Write

\bea 
v\lambda\left( \mathbf{x},\mathbf{x}',\mathbf{x}''\right)&=&\left( 2\pi\right) ^3\int\;\frac{d^3p}{\left( 2\pi\right) ^3}\frac{d^3p'}{\left( 2\pi\right) ^3}\frac{d^3p''}{\left( 2\pi\right) ^3}\;e^{i\left( \mathbf{p} \mathbf{x}+\mathbf{p}' \mathbf{x}'+\mathbf{p}'' \mathbf{x}''\right)  }\nn
&& \delta\left( \mathbf{p}+\mathbf{p}'+ \mathbf{p}''\right) 
f\left(  \mathbf{p}\right) f\left( \mathbf{p}'\right)f\left( \mathbf{p}''\right)
\tea 
then

\bea 
I_B\left[ \mathbf{p},\mathbf{p}'; \mathbf{q}\right]=\omega^6\Delta\epsilon^3\rho f\left(  \mathbf{q}\right) \int\;\frac{d^3r}{\left( 2\pi\right) ^3}&&\left\lbrace G^*\left( \mathbf{r}\right)f\left(  \mathbf{r-p'}\right) f\left( \mathbf{p'-q-r}\right) \right.  \nn
&+&\left.  G\left( \mathbf{r}\right)f\left(  \mathbf{p-r}\right) f\left( \mathbf{q+r-p}\right) \right\rbrace   
\tea
After this analysis, it is straightforward to build the kernel $\mathcal{I}$. We give a diagrammatic representation in figs. (\ref{code}) to (\ref{a3g}). We represent $\mathcal{I}\left[ \mathbf{p},\mathbf{K}\right]$ as a four terminal device, with momentum $\mathbf{p}$ flowing into the upper left vertex and being extracted from the lower left vertex, and momentum $\mathbf{K}$ flowing into the lower right vertex and being extracted from the upper right vertex; momentum conservation holds in internal vertices. We represent $G$ by a right pointing arrow, $G^*$ by a left pointing one, $f$ by a wavy line and $G_1$ by a double arrow. Points indicate that no propagator is assigned to that line. With these conventions (see fig. (\ref{code})) the Bethe-Salpeter equation (\ref{BeSal}) looks like fig. (\ref{besalg}). The different contributions to $\mathcal{I}\left[ \mathbf{p},\mathbf{K}\right]$ are shown in the remaining figures.

When $\mathbf{p}$, $\mathbf{K}\ll R^{-1}$ both $\mathcal{I}_{ladder}$ and $\mathcal{I}_B$ converge to constant values, $\omega^4\Delta\epsilon^2\rho v^2$ and (modulo a numerical factor) $\omega^4\Delta\epsilon^2\rho v^2\left( \omega^2\Delta\epsilon R^2\right) $ respectively. The other two loops contributions are suppressed with respect to  $\mathcal{I}_{ladder}$ and $\mathcal{I}_B$ by a factor of $\rho v$, which we assume is small. However, they remain nontrivial functions of wave number even at long wavelenghts; the maximally crossed graph $\mathcal{I}_A^{\left( 3\right) }$, in particular, is related to the backscattering peak, as can be appreciated from fig. (\ref{a3g}).

\subsection{The self energy}

\begin{figure}[htb]
\begin{center}
\includegraphics[scale=0.6]{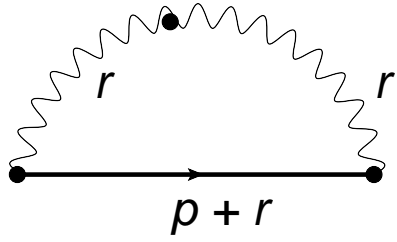}
\caption{The one loop contribution to the self-energy, which reproduces the so-called nonlinear approximation. }
\label{sigma1g}
\end{center}
\end{figure}

\begin{figure}[htb]
\begin{center}
\includegraphics[scale=0.6]{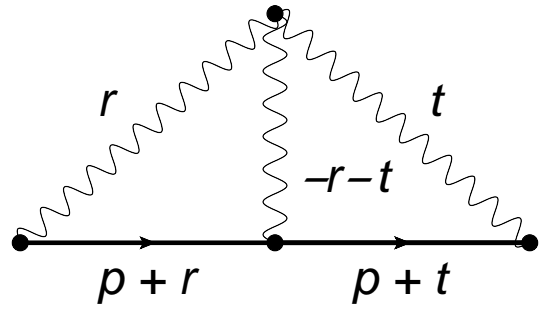}
\caption{The two loops contribution to the self-energy from non gaussian effects. }
\label{sigma2ngg}
\end{center}
\end{figure}

\begin{figure}[htb]
\begin{center}
\includegraphics[scale=0.6]{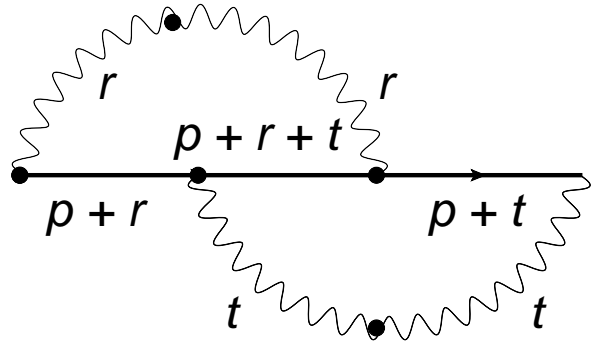}
\caption{The two loops contribution to the self-energy from gaussian correlations. }
\label{sigma2gg}
\end{center}
\end{figure}

When estimating the corrections to the self energy,  we  make the approximation of replacing the self energy by an effective inverse square speed of sound, namely

\be 
\epsilon_{eff}=\bar\epsilon +\frac 1{\omega^2}\int\;dy\;\Sigma\left( x,y\right) 
\te
Under this approximation solving for $G$ is immediate

\be 
G\left( r\right) =\frac{e^{ikr}}r
\te
where $k=\omega\epsilon_{eff}^{1/2}$. We identify three contributions to the self energy

\be 
\Sigma =\Sigma_G^1+\Sigma_{NG}^2+\Sigma_G^2
\te 
The first one is the usual (1-loop) nonlinear approximation eq. (\ref{nonlinear}) (see fig. (\ref{sigma1g}), the second is the 2-loops non gaussian term eq. (\ref{sigma2ng}) (see fig. (\ref{sigma2ngg}), and the third is the 2-loops Gaussian correction eq. (\ref{sigma2g}) (see fig. (\ref{sigma2gg}). We write correspondingly

\be 
\epsilon_{eff}=\bar\epsilon + \epsilon_G^1+\epsilon_G^2+\epsilon_{NG}^2
\te
where 

\be 
\epsilon_G^1=\omega^2\Delta \epsilon^2\rho v\int\;d^3\mathbf{x}\; \frac 1r\gamma\left( \mathbf{x},0\right)=\alpha_G^1R^2\omega^2\Delta \epsilon^2\rho v
\te
$\alpha_G^1\le 8\pi$ is a purely geometrical factor. Similarly

\be 
\epsilon_{NG}^2=\alpha_{NG}^2R^4\omega^4\Delta \epsilon^3 \rho v
\te
where 

\be 
\alpha_{NG}^2=\frac 1{R^4}\int\;d^3\mathbf{x}_1\;d^3\mathbf{x}_2\frac{\lambda\left( \mathbf{x}_1, \mathbf{x}_2,0\right)}{\vert\mathbf{x}_1-\mathbf{x}_2\vert\vert\mathbf{x}_2\vert}
\te 
and

\be 
\epsilon_G^2=\alpha_G^2R^6\omega^6\Delta \epsilon^4\left( \rho v\right) ^2
\te
where 

\be 
\alpha_G^2=\frac 1{R^6}\int\;d^3\mathbf{x}_1\;d^3\mathbf{x}_2\;d^3\mathbf{x}_3\frac{\gamma\left( \mathbf{x}_1,\mathbf{x}_3\right)\gamma\left( \mathbf{x}_2,0\right)}{\vert\mathbf{x}_1-\mathbf{x}_2\vert\vert\mathbf{x}_2-\mathbf{x}_3\vert\vert\mathbf{x}_3\vert}e^{ik\left[ \vert\mathbf{x}_1-\mathbf{x}_2\vert +\vert\mathbf{x}_2-\mathbf{x}_3\vert +\vert\mathbf{x}_3\vert\right] }
\te 
The integrand is zero unless $\vert\mathbf{x}_1-\mathbf{x}_3\vert \le 2R$ and $\vert\mathbf{x}_2\vert\le 2R$. We may then approximate

\be 
\alpha_G^2\approx\int_{\vert\mathbf{x}_3\vert\ge 2R}\;\frac{d^3\mathbf{x}_3}{\vert\mathbf{x}_3\vert^3}e^{3ik\vert\mathbf{x}_3\vert }
\te 
In the limit $R\to 0$ the integral diverges in the lower limit, leading to the estimate

\be 
\alpha_G^2\approx 4\pi\ln{kR}
\te 
Already from the parametric dependence in these estimates we see that there is an ample range of parameters where
$\epsilon_G^1\ge \epsilon_{NG}^2\gg\epsilon_G^2$. Therefore, if the latter is included, e. g., to restore reciprocity without spoiling flux conservation, it would be simply inconsistent not to include the second.

\section{Conclusions}
In this paper we have casted the problem of scalar wave propagation in a random medium in a field theoretic language which connects it immediately to the larger body of work \cite{Zinn93,Kle90} addressed to similar problems in high energy physics and cosmology \cite{CalHu08}  and the theory of turbulence \cite{MonYag71,McC94,Fris95,BelLvov87,Cal09}. The advantages of the method, as compared with the straightforward approach of iterating the Dyson equation,  are  that it provides a partial resummation of the perturbative series which avoids overcounting and has energy conservation built in order by order \cite{PRC09}. As a sample of the power of the method we have analyzed the effect of non gaussian statistics on the mass and intensity operators. We have shown that in a generic model the nongaussian effects are quantitatively dominant over gaussian ones at two loops order. The latter should be considered, however, because only so such effects as backscattering enhancement may be described correctly, but then, for consistence, so should be the former ones.

In last analysis, we believe the final proof of the usefulness of the language we are proposing will be given by its application to the electromagnetic case and the important problem of depolarization. Another feature of the field theory language which may prove decisive is its flexibility towards the application of even more powerful nonperturbative techniques, among which renormalization group improvement of the self energy and intensity operators feature prominently \cite{ZanCal06}. We hope to report shortly on progress in these directions.

\section*{Acknowledgements}

It is a pleasure to thank O. Bruno for a very useful suggestion.

This work is supported in part by CONICET, ANPCyT and Universidad de Buenos Aires (Argentina).

\end{document}